# Mechanism for Embedded In-plane Self Assembled Nanowire Formation

Nathaniel S Wilson,[a] Stephan Kraemer,[c] Daniel J. Pennachio,[a] Patrick Callahan,[a] Mihir Pendharkar,[b] Christopher J Palmstrøm[a,b,+]

[a]Materials Dept. University of California, Santa Barbara 93106-5050, USA

[b]Electrical and Computer Engineering Dept. University of California, Santa Barbara 93106-5050, USA

[c]Center for Nanoscale Systems, Harvard University, Boston, MA 02138, USA

**Abstract:**

We report a novel growth mechanism that produces in-plane $[1\bar{1}0]$ oriented ErSb nanowires formed during codeposition of $Er_{0.3}Ga_{0.7}Sb$ via molecular beam epitaxy (MBE). Nanowires are characterized by in-situ scanning tunneling microscopy (STM), as well as ex-situ transmission electron microscopy (TEM) and electron channeling contrast imaging (ECCI). We show that complexes of macrosteps with step heights on the order of 7 nm form during nanowire growth. The macrosteps are shown to be part of the in-plane nanowire growth process and are directly responsible for the observed stratified distribution of in-plane nanowires. TEM indicates that initial growth results in out-of-plane nanowires transitioning to in-plane nanowires after a critical film thickness. A surface energy model is put forward that shows the critical thickness is due to minimization of the GaSb{110} surfaces formed during out-of-plane nanowire growth. Kinetics of the transition are discussed with respect to observed features in STM, along with the material parameters needed to achieve in-plane nanowire growth.

## I. Introduction

Spontaneous phase separation resulting in self-assembled nanostructures has been observed in a wide range of materials systems including semiconductors,[1] oxides,[2] and metals.[3] . Often these nanostructures show new and desirable properties resulting from orientation,[4] confinement,[5] or increased coupling with a matrix material,[1] making them an exciting area of interest when looking for new material properties. Spinodal decomposition and the resulting microstructures have long been an area of interest in bulk materials[6] but the behavior is less well understood during thin film growth processes such as molecular beam epitaxy (MBE). The role of strain in phase separation during MBE growth has been the subject of many studies[7,8] but the self-assembly of nanostructures in film growth is further complicated by surface specific kinetics and effects. Understanding the processes governing spontaneous phase separation in a surface dominated regime like MBE growth is imperative to enable engineering of self-assembled nanostructures during thin film growth.

The rare-earth/monopnictide (RE-V) – III-V compound semiconductor material system is an ideal test bed for investigating nanostructures formed by spontaneous phase separation. Most RE-Vs compounds have the rocksalt (NaCl) crystal structure and are typically semimetals which were initially studied for use as thermodynamically stable epitaxial contacts to III-V semiconductors.[9–14] Epitaxial growth of RE-V on III-V is relatively easily achieved due to similar lattice constants and a continuous group V atomic lattice between both crystal structures. However, one of the primary challenges for this material is defect formation during overgrowth of III-V's on RE-V due to the difference in the surface symmetry and bonding of the III-V (001) and the RE-V (001) surfaces. They are 2-fold and covalent and 4-fold and ionic for the III-V and RE-V, respectively.[15] Hence, most III-V overgrowth efforts have moved away from complete RE-V thin film coverage to structures with embedded RE-V nanoparticles in a III-V matrix.[16,17] This allows for seeding on a III-V surface and lateral overgrowth over the RE-V, and shows particular promise with the use of surfactants.[18,19] The majority of current research in this area focuses on the ErAs material system where ErAs nanoparticles in a GaAs or InGaAs matrix have found use in a number of applications, such as photomixers,[20] tunnel junctions,[21] thermoelectrics,[3,22] and may be used to influence quantum dot formation.[23] Interest has also recently expanded to similar materials such as TbAs and LuAs, which may form semiconducting nanoparticles,[24] and TbAs/ErAs core-shell nanostructures.[25] All of these applications take advantage of the fact that RE-V materials are well suited to nanoparticle nucleation due to the low solid solubility of rare-earths within III-V semiconductors as well as RE-V stability, resistance to diffusion, compatible crystal structures, and lattice matching.[11,26,27]

The ErSb/GaSb material system studied in this work has exhibited the widest range of different nanostructures seen in a RE-V/III-V material system including isotropic nanoparticles, vertical nanowires, horizontal nanowires, and nanosheets, all of which are accessible by controlling the Er to Ga flux ratio.[28] The formation of nanoparticles and vertical nanowires is explained by an embedded growth mode where Er adatoms replace Ga in the first 3-4 monolayers of the GaSb surface.[28–30] ErSb's lower heat of formation is the driving force behind embedded growth and, combined with the difference in surface energies which discourages GaSb growth on ErSb, results in vertical nanowires.[29] However, this model fails to provide a mechanism for the formation of the stratified layers of horizontal nanowires and nanosheets observed at Er concentrations in excess of 30%. It is important to understand the formation mechanism behind these structures if they are to be incorporated into sensitive device heterostructures and an in-depth understanding of the formation mechanism for these nanostructures may enable identification and engineering of other material systems with similar behavior.

## II. Experimental Procedure

Samples were grown in an arsenic and antimony containing III-V VG-V80H molecular beam epitaxy (MBE) chamber with a background pressure of $<5\times10^{-11}$ Torr. The substrate temperature was measured by an optical pyrometer, calibrated to 540°C at the GaSb(001) substrate oxide desorption temperature as observed by reflection high energy electron diffraction (RHEED). Each sample was grown with a buffer layer of 100-250 nm of GaSb doped with Be at $5\times10^{18}$ cm$^{-3}$ at a substrate temperature of 480°C to ensure a smooth surface before nanowire growth began. For nanowire growth, the substrate temperature was increased to 530°C so that GaSb growth was in a step flow regime.[31] Nanowires were grown by codeposition of Er and Ga from elemental sources along with an overpressure of $Sb_2$ provided by an antimony valved cracker effusion cell. The nanostructures were initially characterized *in-situ* by scanning tunneling microscopy (STM), imaging the growth surface without oxidation or contamination. The STM measurements were performed in an Omicron Low Temperature STM at 77K with a tungsten tip and a tunneling current of 30 pA. Tip bias varied from -2V to 2V depending if the surface was primarily GaSb or ErSb. Scanning tunneling spectroscopy (STS) was performed using a variable tip height method by applying an additional AC voltage on the STM tip and measuring the resulting AC signal with a lock-in amplifier. STS data was processed and normalized to approximate local density of states in the manner described by Feenstra.[32]

## III. Results

An embedded growth mode was confirmed by depositing 0.15 monolayers of Er on a GaSb surface, Figure 1(a). Embedded nucleation is apparent as ErSb has only 3.7% surface coverage instead of 15%, consistent with four monolayer thick nanostructures. The ErSb nanostructures appear to be one monolayer high implying three monolayers are embedded in the GaSb matrix, in agreement with the previously observed embedded growth mode.[28] Figure 1(b) shows typical STS spectra of the sample where a metallic density of states is observed at the ErSb nanoparticles, while the GaSb matrix shows a gapped semiconducting density of states. Identification of surface reconstructions on later samples was performed by comparing STS to the ErSb and GaSb results from this known surface.

## A. Surface Morphology

Previous work has shown that codeposited $Er_xGa_{1-x}Sb$ with Er concentrations of X=0.1 – 0.25 results in ErSb nanowires oriented out-of-plane. These nanowires grow as ErSb preferentially deposits on previously nucleated ErSb particles, elongating the particles in the out-of-plane direction. Figure 2 displays STM topography images for $Er_xGa_{1-x}Sb$ for X=0.2, 0.25, 0.3. The $Er_{0.2}Ga_{0.8}Sb$ surfaces closely match the expected surface for out-of-plane nanowire growth and is characterized by the ends of out-of-plane ErSb nanowires terminating on the surface at the bottom of pits in the GaSb matrix. As the Er content is increased to $Er_{0.25}Ga_{0.75}Sb$ vertical nanowires are still observed although the pits are noticeably larger and deeper. At $Er_{0.3}Ga_{0.7}Sb$ the transition to in-plane nanowires has occurred and pits around the nanowires have disappeared resulting in what appears to be a much smoother surface with ErSb nanowires lying in plane with the GaSb(001) surface Figure 2(c). Clearly there has been a dramatic change in the surface morphology for x > 0.25. The regions corresponding to GaSb and ErSb are clearly distinguishable in the STM image by the different surface reconstructions in Figure 2(c). The checker board like reconstruction and small islands seen on the ErSb is consistent with reconstructions observed previously for 10 monolayer ErSb thin films.[29] Both the checker board and small island reconstructed regions exhibit metallic density of states when probed by STS. Values of ErSb surface coverage reported consist of the total combined surface coverage of both checker board and small island surface reconstructions associated with ErSb.

Although on the atomic scale, the surface looks smooth for x=0.3, macrosteps on the order of 7 nm tall are observed on the surface at a larger scale as seen in the scanning electron microscope (SEM) image of Figure 3(a). The spacing between macrosteps appears to vary with typical values being on the order of 500-1000 nm.

Figure 3(b-f) shows the STM topography along a terrace and across a macrostep. From the relative changes of the coverage of the different surface reconstructions, a change in ErSb concentration based on position within the terrace is clearly evident. Areas near the bottom of a macrostep Figure 3(d) exhibit very large concentrations of ErSb (88%) while areas near the top edge of a macrostep Figure 3(c) show a much lower ErSb surface coverage of 14%. This suggests that there must be substantial diffusion of either Er or Ga on the terrace surface. Previous results of ErSb growth on GaSb observed an Er-Ga exchange reaction and kick-out of Ga that diffused across the surface to grow GaSb at step edges.[28] Hence, it is more likely that Ga is the faster diffusing species on the terrace surface. Note further that right on the top edge of a step, the GaSb coverage looks nearly continuous Figure 3(b). The shape of the GaSb fingers going back from the step edge is suggestive of GaSb step flow growth back on the terrace from the step edge. This would be consistent with diffusing Ga on the terrace surface being reflected back by the step edge, as would be expected from a Ehrlich-Schwoebel barrier at the step edge.[33] The overall surface morphology can be explained by stepflow of the macrosteps resulting from Ga not wetting the terrace surface and diffusion on the terrace surface to the macrostep edges during growth. Surfaces near the top edges of macrosteps are relatively young as the macrostep, being formed by both lateral surface diffusion of Ga and Ga deposition, overgrows the area beneath. Areas near the base of the macrostep have been exposed to Er adatom deposition for longer periods of time and are about to be overgrown. This results in the large difference of observed ErSb coverage.

## B. Macrostep structure

Cross-sectional high angle annular dark field scanning transmission electron microscopy (HAADF-STEM) was performed on an additional $Er_{0.3}Ga_{0.7}Sb$ sample grown under identical conditions to the sample shown in Figure 4. This sample had ~25 nm of amorphous $AlO_x$ deposited in-situ via e-beam evaporation to preserve macrostep features on the surface. Figure 4(a) and 4(c) show a macrostep looking along the [110] and [1$\bar{1}$0] zone axes respectively. When viewed along these directions the zincblende crystal has a higher projected areal atomic density, which, in combination with the higher average atomic number (Z) of ErSb, makes the nanoparticles easily distinguishable as the brighter areas due to the Z contrast of HAADF-STEM. Macrosteps oriented along the [110] and [1$\bar{1}$0] directions both show a predominantly uninterrupted nanowire containing layer underneath them indicating that macrosteps grow in a step-flow fashion across the surface. In contrast, a stationary macrostep should coincide with a discontinuity in the underlying nanowire layers.

Figure 4(b) and 4(d) show the first observable nanoparticle on top of the macrostep. For the [1$\bar{1}$0] facing macrostep (Figure 4(a)) the ErSb is 100nm from the step edge while the [110] facing macrostep (Figure 4(b)) shows nanoparticles within 15 nm of the step edge. This discrepancy in macrostep growth is attributed to the asymmetric diffusion rates along the GaSb(001) surface[29] and is in agreement with the observed asymmetry in the nanowires and macrosteps observed in Figure 3.

Overgrowth of GaAs on planar ErAs layers has been shown to result in nucleation of twinned GaAs islands.[34] To investigate the possibility of similar behavior in ErGaSb growth a 25 nm thick $Er_{0.3}Ga_{0.7}Sb$ sample was grown without an $AlO_x$ capping layer for characterization by ECCI. ECCI is sensitive to atomic displacements in a crystal such as the displacement from strain fields in dislocations and should reveal the presence of defects driving nucleation. Figure 5 shows an ECCI image of the surface of an $Er_{0.3}Ga_{0.7}Sb$ film. A misfit dislocation can be seen as a white line rising up to the surface at the top of the image. A few other misfit dislocations were observed using ECCI but no other defects were seen, implying there is no defect present at a density similar to that of the macrosteps on the surface responsible for their formation.

One benefit of the ECCI measurement is that the backscattered electrons in the measurement clearly show the high Z ErSb nanowires. Figure 6 shows SEM images taken in the ECCI channeling conditions, backscatter mode, and secondary electron mode. Since ErSb is a very similar crystal structure to GaSb many of the channeling conditions work for both phases. As a result, the ECCI image only shows ErSb nanowires directly on the surface where incident electrons behave as they would in a normal backscatter image. Deeper nanowires are invisible due to the presence of the channeling effect of the atomic lattice. In contrast the backscatter image shows all nanowires throughout the 25nm thick layer, the secondary electron image primarily shows the surface morphology of the macrostep. The ECCI image in Figure 6(a) clearly shows that the surface nanowires near areas of recent growth such as the macrostep edge and the notch feature in the top right are smaller and less dense enforcing the conclusion from STM and TEM that macrostep flow occurs during growth, and nanowire coverage is a function of Ga, Er, and Sb flux exposure time.

## IV. Surface Energy Model

Previously the nanowire surface energy was given as a possible explanation for the driving force responsible for the changes in nanowire morphology as a function of Er concentration.[28] However, this hypothesis fails to explain the initial growth of $Er_{0.3}Ga_{0.7}Sb$ consisting of out-of-plane nanowires with horizontal growth beginning after the first layer is terminated by a macrostep. We postulate that it is not the surface energy of the nanowires driving the transition but instead the total surface area of the GaSb{110} surface which consists of the walls of the pits observed in the STM images. This driving force can be easily explained by a simple model which compares the surface energy of the GaSb in the vertical nanowire (VNW) growth regime, where excess GaSb grows around nanoparticles creating a porous structure with GaSb{110} sidewalls, and horizontal nanowires, where the excess GaSb which previously formed the pits around the nanoparticles is condensed into a large rectangular macrostep on the surface, a schematic of these two surface morphologies is shown in Figure 7**Error! Reference source not found.**(a).

The total surface energy of a surface is given by

(1) $E = \sum S_{GaSb\{110\}} \gamma_{GaSb\{110\}} + \sum S_{GaSb(001)} \gamma_{GaSb(001)} + \sum S_{ErSb(001)} \gamma_{ErSb(001)} + \sum S_{\frac{ErSb}{GaSb(001)}} \gamma_{\frac{ErSb}{GaSb(001)}}$

The $S_{GaSb\{110\}}$, $S_{GaSb\{001\}}$, $S_{ErSb(001)}$, $S_{ErSb/GaSb(001)}$, terms in equation (1) represent the surface areas of the GaSb{110}/vacuum, GaSb(001)/vacuum, ErSb(001)/vacuum surfaces and the buried GaSb(001)/ErSb(001) interface formed when a macrostep overgrows a nanowire, the $\gamma_{GaSb\{110\}}$, $\gamma_{GaSb(001)}$, $\gamma_{ErSb(001)}$, $\gamma_{ErSb/GaSb(001)}$, terms represent the surface energies of the GaSb{110}/vacuum, GaSb(001)/vacuum, ErSb(001)/vacuum surfaces and the buried GaSb(001)/ErSb(001) interface. The buried interfaces between the ErSb nanowires and GaSb matrix not explicitly shown in equation 1 are assumed to have equivalent surface areas in both horizontal and vertical nanowire morphologies, this is true if the shape of a nanowire is the same in each case. In this model it is assumed that the surface area coverage of GaSb(001) and ErSb(001) is unchanged directly before and after the transition occurs, and that when calculating ΔE, the difference between the total surface energy of the in-plane nanowires ($E_{\parallel}$) and the total surface energy of the vertical nanowires ($E_{\perp}$) cancel resulting in equation (2).

(2) $\Delta E = E_{\parallel} - E_{\perp} = \sum S_{GaSb\{110\}} \gamma_{GaSb\{110\}} + \sum S_{\frac{ErSb}{GaSb\{001\}}} \gamma_{\frac{ErSb}{GaSb\{001\}}} - \sum S'_{GaSb\{110\}} \gamma_{GaSb\{110\}}$

Where the $S_{GaSb\{110\}}$ and $S_{ErSb/GaSb\{001\}}$ terms represent the in-plane nanowire growth surface and the $S'_{GaSb\{110\}}$ represents the out-of-plane growth surface. The transition between surfaces occurs when ΔE = 0 which results in the ratio shown in equation 3.

(3) $\frac{E_{\parallel}}{E_{\perp}} = 1 = \frac{\sum S_{GaSb\{110\}} \gamma_{GaSb\{110\}} + \sum S_{\frac{ErSb}{GaSb\{001\}}} \gamma_{\frac{ErSb}{GaSb\{001\}}}}{\sum S'_{GaSb\{110\}} \gamma_{GaSb\{110\}}}$.

Substituting for particle and macrostep dimensions, assuming square-based pits around vertical nanowires and square macrostep features, results in the expression

(4) $\frac{E_{\parallel}}{E_{\perp}} = C_1 \frac{\sqrt{(1-x)}}{x} + C_2(1-x)$.

$C_1 = \frac{h_{ms} l_{pit} \sqrt{\frac{A h_{pit}}{h_{ms}}}}{h_{pit} A}$, $C_2 = \frac{\gamma_{GaSb/ErSb}}{\gamma_{GaSb\{110\}}} \frac{l_{pit}}{4 h_{ms}}$,

Where $h_{ms}$ is the height of the macrostep, $l_{pit}$ and $h_{pit}$ are the base length and height of a vertical nanowire pit, x is the Er content where x = 1 is pure ErSb and *A* is the total surface area contributing to macrostep formation. This model is plotted in Figure 7, with values of $h_{ms}$=7nm, $l_{pit}$=4nm, and $h_{pit}$=1.5nm as estimated from STM measurements. The surface energy component of $C_2$ is unknown, but merely shifts the position of the HNW/VNW crossover without impacting the overall shape of the plot.

As such, a value of $\frac{\gamma_{GaSb/ErSb}}{\gamma_{GaSb\{110\}}} = 8$ was used to provide a crossover similar to what is observed in experiment. Several different values of A can be seen plotted in Figure 7. A represents the area of the surface contributing GaSb to a single macrostep and would vary based on the temperature-dependent diffusion length of Ga on the surface. The model predicts that the Er content required to reach the critical threshold for macrostep formation will monotonically decrease as substrate growth temperature increases. This model provides a clear driving force for macrostep formation motivated from the better surface to volume ratio achieved for GaSb{110} surfaces at higher Er concentrations.

## V. Proposed HNW formation mechanism

There remain several surface features that are not covered in the described model implying that the surface structure is heavily influenced by kinetics, as would be expected since MBE growth is not an equilibrium process. The model assumes a constant height of 7nm for the macrosteps based on the STM images, however, the GaSb(001) surface has a greater surface energy than the GaSb{110} or GaSb{111} indicating that either taller macrosteps or some form of faceting would be more energetically favorable if surface energies were the only consideration.[36] Other features which may be explainable by kinetics include macrosteps forming more complex structures than large isotropic islands on the growth surface, and the elongation of nanoparticles to horizontal nanowires.

Horizontal nanowire elongation can be explained if the surface energy of the ErSb{110}/vacuum interface is much larger than the ErSb(001)/vacuum interface and rather than extending a vertical nanowire above the GaSb plane the ErSb instead prefers to diffuse to the edge of a nanowire and embed itself in the GaSb matrix. This situation only arises when macrostep formation has removed the pits that surround ErSb particles at lower Er content. The elongation of nanoparticles can be explained by the asymmetric diffusion rates of the c(2x6) reconstructed GaSb(001) surface which has an easy diffusion direction down the dimer rows along the $[1\bar{1}0]$, a feature seen on similar surface reconstructions in GaAs.[37]

Macrostep morphology is more difficult to explain but may be related to changes in step flow growth brought about by ErSb particles lying in plane with the surface. Typically during stepflow growth diffusion rates are high enough that no attachment on the terrace occurs, Ga adatoms will diffuse all the way to a step edge where bonding is more favorable before bonding. The type of step-flow growth can be characterized by the probability of an adatom attaching to an upstep versus a downstep: when the upstep is more favored terrace growth is driven towards equally spaced terraces and is the growth mode seen in step flow growth of pure GaSb, but if the down step is favored, large terraces outgrow smaller terraces and step bunching occurs.[38] Here it is proposed that a shift in attachment rate at step edges occurs during ErSb growth due to the presence of dense ErSb coverage observed at the base of macrosteps. The critical ErSb coverage occurs for x ≥0.3. This can be considered schematically in Figure 8(a). In this growth model Ga adatoms do not adhere at the base of a macrostep due to the ErSb present and growth instead occurs at the downstep of a macrostep. Evidence of this can be seen in Figure 3(b), from the STM images, the consequences of a Ehrlich-Schowebel barrier are observed at the downstep.[39]. The Ehrlich-Schowebel barrier is responsible for the monolayer steps growing backwards from the downstep. No similar monolayer step growth is observed from the base of the macrostep indicating that Ga adatoms are almost completely reflected at the macrostep base and that growth must occur almost entirely as macrostep flow with a small amount of Ga adatoms depositing on the terrace surface near the top macrostep edge due to the Ehrlich-Schowebel barrier. Figure 8(b) shows a ball-and-stick model of the unfavorable atomic bonding that would occur as a GaSb step overgrows an ErSb nanoparticle.[39,40]

Some similarities to this suggested macrostep growth mode are observed in other material systems, most notably the growth of $Si_{.985}C_{.015}$ by MBE on a (118) Si substrate.[41] Kinetic Monte Carlo simulations predict step bunching to occur when an added impurity reduces the binding energy of the adatoms thereby increasing their diffusivity,[42] an interaction observed here due to Ga adatoms preference to avoid nucleating on ErSb particles. The large difference in the observed threshold concentration needed to form macrosteps, 1.5% C in Si[41] and 30% Er in GaSb may be explained by the presence of ErSb in pits during vertical nanowire growth figure 2(a), where the pits counteract the increase in diffusivity the ErSb would otherwise generate. Once nanoparticles lie in plane with the surface the change in surface diffusivity further favors macrostep growth. The beginning of the transition to horizontal nanowires can be seen in Figure 9. Observed step bunching may indicate the beginning of macrostep formation. Simultaneously, ErSb particles are observed just one monolayer below the growth surface, instead of recessed in pits, as expected from the proposed mechanism.

In conclusion we show that the transition from vertical to horizontal nanowires in the ErSb/GaSb system is the result of a new growth mode brought about by a change in surface morphology, characterized by the presence of large macrosteps and in-plane rather than recessed ErSb nanoparticles. The growth of horizontal nanowires is due to the asymmetric diffusion rates of Er on the GaSb (001) surface resulting in preferential elongation along the $[1\bar{1}0]$ direction while vertical structures are prevented from forming due to macrostep growth. The formation of macrosteps is shown to be energetically favorable upon reaching a critical composition of Er, and their presence explains the observed stratification of horizontal nanowires observed in cross-sectional STEM. Knowledge of this macrostep mediated growth mode implies that care must be taken when integrating horizontal nanowires into device heterostructures especially near sensitive layers. Finally the underlying mechanism for formation of horizontal nanowires observed here is potentially applicable to any pair of materials with similar diffusion and surface energy properties: further RE-V/III-V pairs are obvious candidates but other nanostructures may display similarities as well.

**Acknowledgements**

This work was funded by the National Science Foundation DMR-1507875, we also acknowledge the use of facilities within the National Science Foundation Materials Research and Science and Engineering Center (DMR 11–21053) at the University of California: Santa Barbara.

[1] D. Jung, J. Faucher, S. Mukherjee, A. Akey, D.J. Ironside, M. Cabral, X. Sang, J. Lebeau, S.R. Bank, T. Buonassisi, O. Moutanabbir, and M.L. Lee, Nat. Commun. **8**, 1 (2017).

[2] J. Hemberger, A. Krimmel, T. Kurz, H.-A. Krug von Nidda, V.Y. Ivanov, A.A. Mukhin, A.M. Balbashov, and A. Loidl, Phys. Rev. B **66**, 94410 (2002).

[3] H. Lu, P.G. Burke, A.C. Gossard, G. Zeng, A.T. Ramu, J.H. Bahk, and J.E. Bowers, Adv. Mater. **23**, 2377 (2011).

[4] H. Lu, D.G. Ouellette, S. Preu, J.D. Watts, B. Zaks, P.G. Burke, M.S. Sherwin, and A.C. Gossard, Nano Lett. **14**, 1107 (2014).

[5] D. Leonard, M. Krishnamurthy, C.M. Reaves, S.P. Denbaars, and P.M. Petroff, Appl. Phys. Lett. **63**, 3203 (1993).

[6] J.W. Cahn, Acta Metall. **9**, 795 (1961).

[7] F. Léonard and R.C. Desai, Phys. Rev. B **57**, 4805 (1998).

[8] F. Léonard, M. Laradji, and R.C. Desai, Phys. Rev. B **55**, 1887 (1997).

[9] C.J. Palmstrøm, N. Tabatabaie, and S.J. Allen, Appl. Phys. Lett. **53**, 2608 (1988).

[10] D.C. Driscoll, M.P. Hanson, E. Mueller, and A.C. Gossard, J. Cryst. Growth **251**, 243 (2003).

[11] A. Guivarc'h, A. Le Corre, P. Auvray, B. Guenais, J. Caulet, Y. Ballini, R. Gúcrin, S. Députier, M.C. Le Clanche, G. Jézéquel, B. Lépine, A. Quémerais, and D. Sébilleau, J. Mater. Res. **10**, 1942 (1995).

[12] E.M. Krivoy, H.P. Nair, A.M. Crook, S. Rahimi, S.J. Maddox, R. Salas, D.A. Ferrer, V.D. Dasika, D. Akinwande, and S.R. Bank, Appl. Phys. Lett. **101**, (2012).

[13] A.G. Petukhov, W.R.L. Lambrecht, and B. Segall, Phys. Rev. B **50**, 7800 (1994).

[14] W.R.L. Lambrecht, B. Segall, A.G. Petukhov, R. Bogaerts, and F. Herlach, Phys. Rev. B **55**, 9239 (1997).

[15] C. Kadow, J.A. Johnson, K. Kolstad, J.P. Ibbetson, and A.C. Gossard, J. Vac. Sci. Technol. B **18**, 2197 (2000).

[16] C.C. Bomberger, M.R. Lewis, L.R. Vanderhoef, M.F. Doty, J.M.O. Zide, J. Vac. Sci. Technol. B **35**, 30801 (2017).

[17] B.E. Tew, M.R. Lewis, C. Hsu, C. Ni, and J.M.O. Zide, J. Cryst. Growth **518**, 34 (2019).

[18] R. Salas, S. Guchhait, K.M. McNicholas, S.D. Sifferman, V.D. Dasika, D. Jung, E.M. Krivoy, M.L. Lee, and S.R. Bank, Appl. Phys. Lett. **108**, 182102 (2016).

[19] R. Salas, S. Guchhait, S.D. Sifferman, K.M. McNicholas, V.D. Dasika, D. Jung, E.M. Krivoy, M.L. Lee, and S.R. Bank, APL Mater. **5**, 96106 (2017).

[20] C. Kadow, A.W. Jackson, A.C. Gossard, J.E. Bowers, S. Matsuura, and G.A. Blake, Phys. E Low-Dimensional Syst. Nanostructures **7**, 97 (2000).

[21] H.P. Nair, A.M. Crook, and S.R. Bank, Appl. Phys. Lett. **96**, 222104 (2010).

[22] R. Koltun, J.L. Hall, T.E. Mates, J.E. Bowers, B.D. Schultz, and C.J. Palmstrøm, J. Vac. Sci. Technol. B Microelectron. Nanom. Struct. **31**, 41401 (2013).

[23] Y. Zhang, K.G. Eyink, L. Grazulis, M. Hill, J. Peoples, and K. Mahalingam, J. Cryst. Growth 477 19 (2017).

[24] C.C. Bomberger, L.R. Vanderhoef, A. Rahman, D. Shah, D.B. Chase, A.J. Taylor, A.K. Azad, M.F. Doty, and J.M.O. Zide, Appl. Phys. Lett. **107**, 0 (2015).

[25] P. Dongmo, M. Hartshorne, T. Cristiani, M.L. Jablonski, C. Bomberger, D. Isheim, D.N. Seidman, M.L. Taheri, and J. Zide, Small **10**, 4920 (2014).

[26] P.F. Miceli, C.J. Palmstrøm, and K.W. Moyers, Appl. Phys. Lett. **58**, 1602 (1991).

[27] A.J. Young, B.D. Schultz, and C.J. Palmstrøm, Appl. Phys. Lett. **104**, 73114 (2014).

[28] J.K. Kawasaki, B.D. Schultz, H. Lu, A.C. Gossard, and C.J. Palmstrøm, Nano Lett. **13**, 2895 (2013).

[29] B.D. Schultz, S.G. Choi, and C.J. Palmstrøm, Appl. Phys. Lett. **88**, 243117 (2006).

[30] B.D. Schultz and C.J. Palmstrøm, Phys. Rev. B - Condens. Matter Mater. Phys. **73**, 1 (2006).

[31] S. Chalmers, H. Kroemer, and A. Gossard, Appl. Phys. Lett. **57**, 1751 (1990).

[32] R.M. Feenstra, Phys. Rev. B **50**, 4561 (1994).

[33] R.L. Schwoebel and E.J. Shipsey, J. Appl. Phys. **37**, 3682 (1966).

[34] T. Sands, C.J. Palmstrøm, J.P. Harbison, V.G. Keramidas, N. Tabatabaie, T.L. Cheeks, R. Ramesh, and Y. Silberberg, Mater. Sci. Reports **5**, 99 (1990).

[35] M. Volmer and W. Schultze, Z. Phys. Chem. **156**, 1 (1931).

[36] J.W. Cahn and R.E. Hanneman, Surf. Sci. **1**, 387 (1964).

[37] E. Penev, S. Stojković, P. Kratzer, and M. Scheffler, Phys. Rev. B **69**, 115335 (2004).

[38] J.Y. Tsao, *Materials Fundamentals of Molecular Beam Epitaxy* (Academic Press, Inc., San Diego, 1993).

[39] P.F. Miceli and C.J. Palmstrøm, Phys. Rev. B **51**, 5506 (1995).

[40] K. Momma and F. Izumi, *J. Appl. Crystallogr.*, **44**, 1272-1276 (2011).

[41] E.T. Croke, F. Grosse, J.J. Vajo, M.F. Gyure, M. Floyd, and D.J. Smith, Appl. Phys. Lett. **77**, 1310 (2000).

[42] J. Vollmer, J. Hegedüs, F. Grosse, and J. Krug, New J. Phys. **10**, (2008).

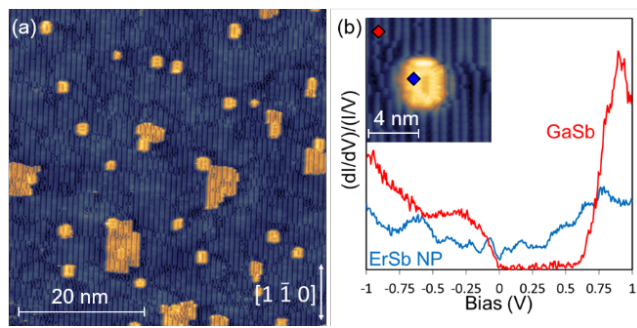

Figure 1: (a) STM of 0.15 monolayers of ErSb deposited on GaSb(001) surface, -0.3V bias. (b) STS of ErSb nanoparticle where GaSb (red) shows a clear gap in its density of states, and ErSb (blue) appears metallic.

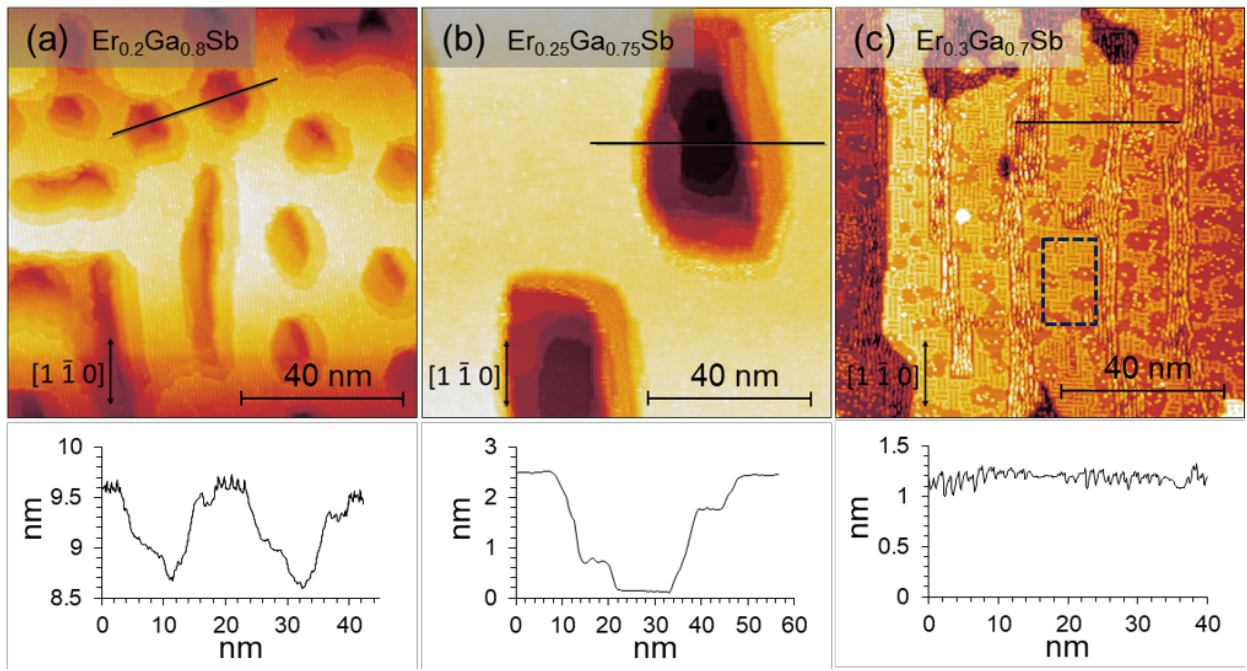

Figure 2: STM and line scan of (a) $Er_{0.2}Ga_{0.8}Sb(001)$ surface with ErSb particles at base of pits, (b) $Er_{0.25}Ga_{0.75}Sb(001)$ surface with ErSb particles at base of pits and (c) $Er_{0.3}Ga_{0.7}Sb(001)$ surface where ErSb nanowires lie in the surface plane, the dashed rectangle indicates a region of ErSb. Line scans in each case are across an ErSb nanoparticle.

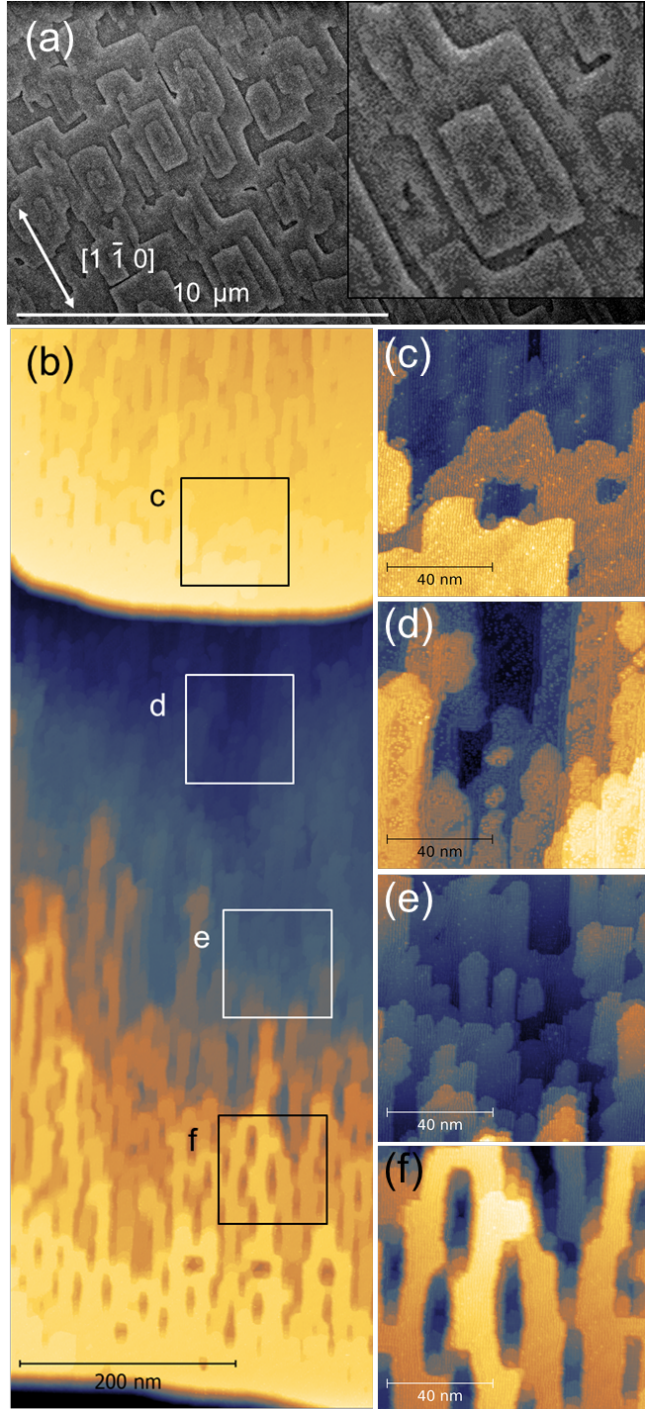

Figure 3: (a) SEM of $Er_{0.3}Ga_{0.7}Sb(001)$ surface, inset shows a blowup of a spiral macrostep feature. (b) STM depicting a pair of macrosteps and the changing surface morphology across the terrace between them. Highlighted areas show positions of higher resolution images used to calculate ErSb surface coverage: (c) 14% ErSb, (d) 88% ErSb, (e) 70% ErSb, (f) 5% ErSb.

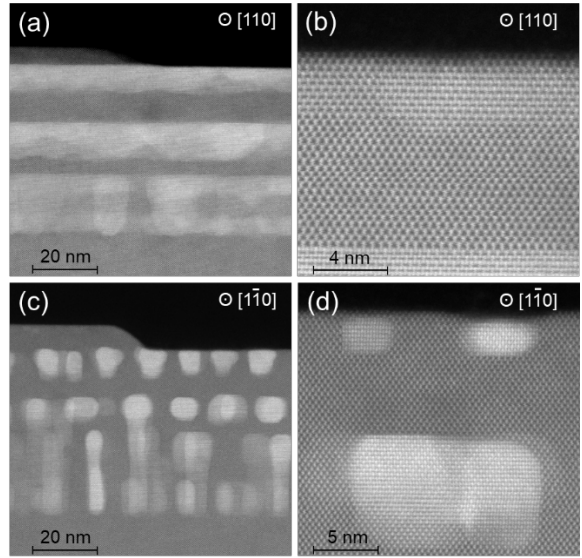

Figure 4: HAADF-STEM of $Er_{0.3}Ga_{0.7}Sb$ samples capped with amorphous $AlO_x$. Lighter regions correspond to ErSb nanowires. (a) Image along the [110] direction perpendicular to the long axis of the nanowires. (b) First observable nanoparticles on the surface of the macrostep in (a) ~100nm from the macrostep edge. (c) Image along the [1$\bar{1}$0] direction parallel long axis of nanowires. (d) First observable nanoparticles on the macrostep seen in (c) ~15nm from step edge.

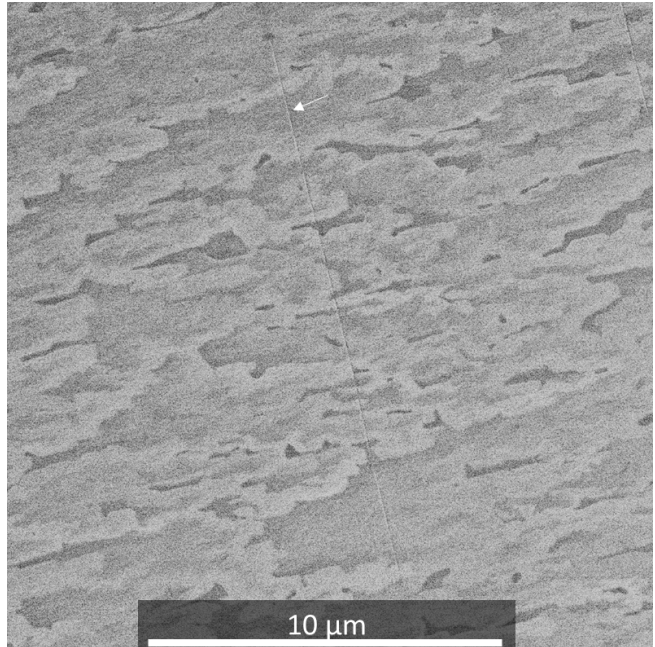

*Figure 5: ECCI image of a 25nm Er$_{0.3}$Ga$_{0.7}$Sb film grown on a GaSb(001) substrate. The only defects with observable strain fields in the sample are misfit dislocations, as seen in the center of the image. The spiral macrostep structure is not observed on this surface.*

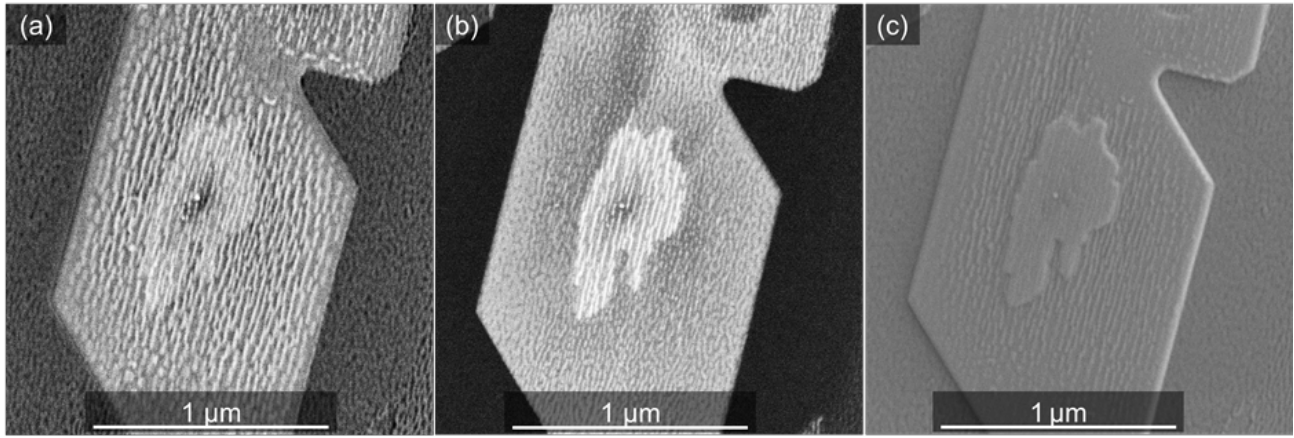

*Figure 6: A macrostep on a 25nm $Er_{0.3}Ga_{0.7}Sb(001)$ surface imaged in three different modes, (a) ECCI showing surface ErSb nanowire distribution. (b) Backscatter electron image tilted 2.5° out of the channeling condition used for Figure (a), nanowires from layers beneath the surface are clearly visible. (c) Secondary electron image showing surface features of the macrostep.*

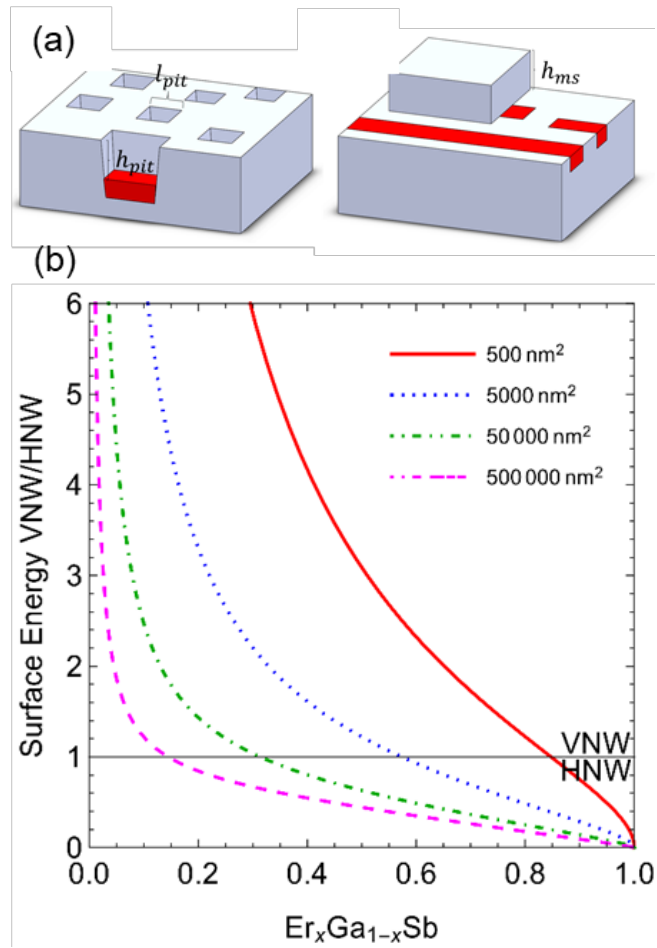

*Figure 7:* (a) Diagram showing the two surface configurations used in the model. The macrostep on the right is comprised of the GaSb above the nanoparticles ($h_{pit}$). (b) Plot of Equation 2 showing the ratio of surface energies between vertical nanowire and horizontal nanowire morphologies. Lines depict different sized areas contributing to a single macrostep. The transition occurs at lower Er concentrations as surface area of a single macrostep increases.

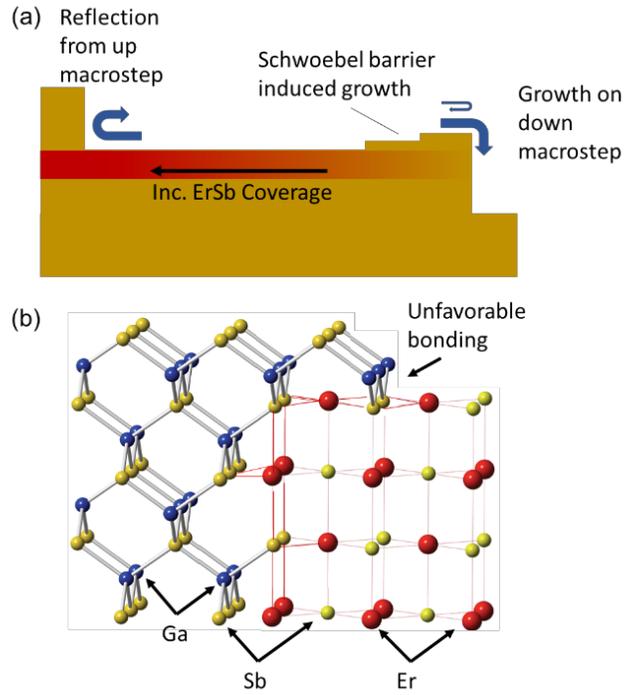

*Figure 8: (a) Diagram showing the growth model for Ga adatoms diffusing along the surface. Ga adatoms reflect from the bottom of a macrostep such that the majority of growth occurs from the top edge of a macrostep, while some growth occurs on the leading edge of the macrostep due to the Schwoebel barrier. (b) Ball-and-stick model showing overgrowth of an ErSb nanoparticle by GaSb.*

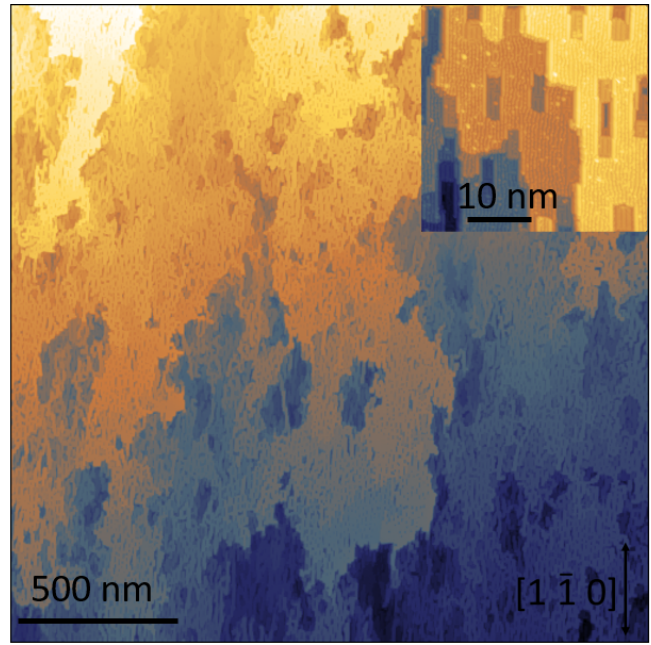

*Figure 9:* STM of 7.5 nm of Er$_{0.3}$Ga$_{0.7}$Sb(001) surface, the beginning of step bunching can be seen. Inset shows ErSb nanoparticles are present one monolayer below the growth plane.